\documentclass[letterpaper, 10 pt, conference]{ieeeconf}  
\usepackage{amsmath}
\usepackage{subcaption}
\usepackage{graphicx}
\usepackage{authblk}
\IEEEoverridecommandlockouts                              

\title{\LARGE \bf
RPnet: A Deep Learning approach for robust R Peak detection in noisy ECG
}

\author{Sricharan Vijayarangan$^{1}$  , Vignesh R$^{1*}$,  Balamurali Murugesan$^{1,2*}$, Preejith SP$^{1}$ \\ Jayaraj Joseph$^{1}$ and  Mohansankar Sivaprakasam$^{1,2}$  
\thanks{* Equal Contribution}
\thanks{$^{1}$ are with Healthcare Technology and Innovation Center (HTIC),
        Indian Institute of Technology (IIT-M), India
        {\tt\small sricharanv@htic.iitm.ac.in}}%
\thanks{$^{2}$ is with Department of Electrical Engineering,
        Indian Institute of Technology, Madras (IITM), India
        {}}%
}
\begin{document}
\maketitle
\thispagestyle{empty}
\pagestyle{empty}
\begin{abstract}
Automatic detection of R-peaks in an Electrocardiogram signal is crucial in a multitude of applications including Heart Rate Variability (HRV) analysis and Cardio Vascular Disease(CVD) diagnosis. Although there have been numerous approaches that have successfully addressed the problem, there has been a notable dip in the performance of these existing detectors on ECG episodes that contain noise and HRV Irregulates. On the other hand, Deep Learning(DL) based methods have shown to be adept at modelling data that contain noise. In image to image translation, Unet is the fundamental block in many of the networks. In this work, a novel application of the Unet combined with Inception and Residual blocks is proposed to perform the extraction of R-peaks from an ECG. Furthermore, the problem formulation also robustly deals with issues of variability and sparsity of ECG R-peaks. The proposed network was trained on a database containing ECG episodes that have CVD and was tested against three traditional ECG detectors on a validation set. The model achieved an F1 score of 0.9837, which is a substantial improvement over the other beat detectors. Furthermore, the model was also evaluated on three other databases. The proposed network achieved high F1 scores across all datasets which established its generalizing capacity. Additionally, a thorough analysis of the model's performance in the presence of different levels of noise was carried out.                    

\end{abstract}
\section{INTRODUCTION}
The Electrocardiogram (ECG) has been one of the most significant elements in the non-invasive diagnosis of cardiac ailments. Accurate detection of R-peaks in an ECG serves as a precursor to the diagnosis of numerous cardiac disease and Heart Rate variability (HRV) analysis. Initially, R-peak detection involved a simple combination of digital filters. Since then, there have been multiple attempts at improving the accuracy of the beat detection algorithm and its computational efficiency. Some of them include wavelet transform, empirical mode decomposition (EMD), neural networks and hybrid approaches \cite{manikandan2012novel}. Most of the mentioned methods evaluate their detectors, exclusively on the MIT-BIH arrhythmia database \cite{moody2001impact} which contains high quality ECG signals. However, accurate R-peak detection is still challenging in the presence of noise which includes signals with low SNR and HRV irregulates (arrhythmias and motion artifacts). This has been verified in a recent study \cite{liu2018performance}, that highlights the failure of top QRS detectors when tested on a noisy ECG database. In fact, they report that none of the popular beat detectors exceed a F1 score of 80 percent when tested on noisy ECG episodes. The aforementioned detectors , which use traditional signal processing methods, are unable to model the noisy segments of the ECG.

Deep learning(DL) based modeling methods have been used to solve a wide range of problems in Computer Vision and Natural Language Processing, providing state of the art results across multiple domains. \cite{sannino2018deep} and \cite{andersen2019deep} use DL to model ECG's in order to identify different arrhythmia types. However, these methods use traditional signal processing to segment R-peaks in the preprocessing stage, which may not work well under noisy conditions. 

The number of R-peaks in an ECG of fixed length is variable and sparse making it challenging to formulate the problem as a straight forward regression task. To address the issue, this work draws inspiration from image segmentation which exploits a Distance Transform (DT) representation of the mask\cite{murugesan2019psi}. The DL model is thus entasked to obtain a DT representation  of the R-peaks from the ECG signal. The R-peaks can then be extracted from this DT representation, with relative ease. In summary, the contribution of the paper are as follows:
\begin{enumerate}
   \item We propose a novel application of the IncRes-Unet to produce a distance map, from which the R-peaks in the ECG can be obtained through minimal post-processing.    
   \item We quantitatively evaluate the performance of the proposed model relative to three other baselines. 
   \item We establish the generalization capacity of the proposed model by evaluating it across three different databases.
   \item We analyze the performance of the proposed model under varying levels of SNR and establish the robustness of our model at very low SNR's.
\end{enumerate}

\section{METHODOLOGY}
\subsection{Problem Formulation}
The problem is framed as a regression task that involves obtaining the Distance Transform (DT) from the input ECG signal. In imaging, the DT map conveys the distance of a particular pixel in the foreground to the boundary between the foreground and the background. In our case, the R-peaks in an ECG are considered analogous to the boundary in images. Here, the DT map
conveys the distance of each point on the ECG from the nearest R-peak. Subsequently, the R-peak locations are easily obtained from the valleys in the DT. The dataset $X = { (x^{(1)},y^{(1)}),(x^{(2)},y^{(2)}),....,(x^{(m)},y^{(m)}) }$, consists of the input ECG signal $x^{(i)}$ and distance map of the ECG $y^{(i)}$, where ($x^{(i)},y^{(i)} \in R^{n}$). Note that $y^{(i)}$ was obtained from the R peak locations. The reference DT $y^{(i)}$ is of the same size as the input ECG signal $x^{(i)}$.

\begin{figure}[t]
    \centering
    \includegraphics[width=\linewidth]{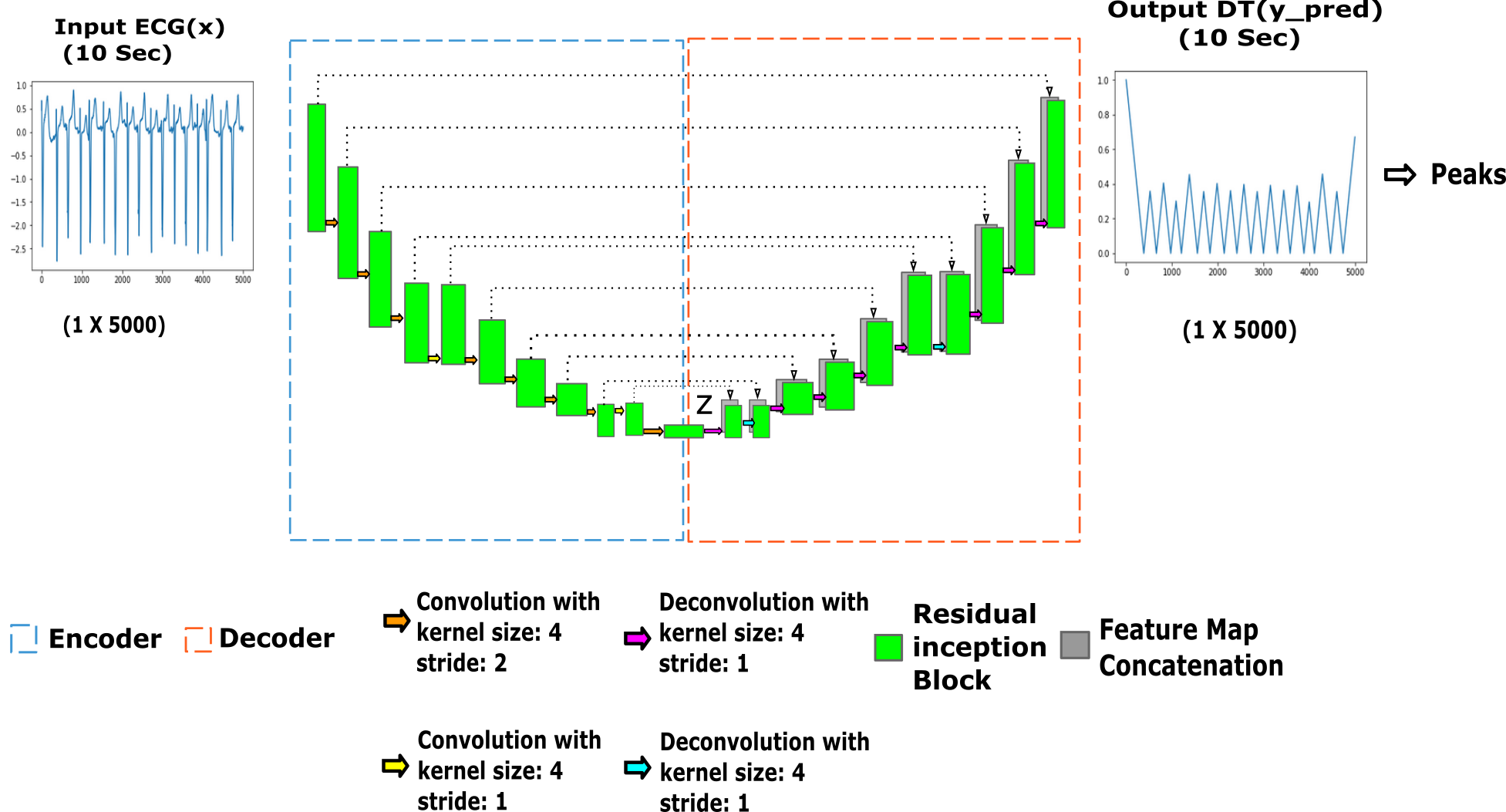}
    \caption{Architecture Details}
    \label{fig:Architecture}
\end{figure}

The proposed network employs a fully convolutional, Encoder-Decoder architecture. The encoder section takes $x^{(i)}$ as input and subsequently downsamples it multiple times to produce the compressed feature vector $Z^{(i)}$. The decoder, upsamples the feature vector obtained from the bottleneck layer to produce the predicted output $y_{pred}$. Equations 1 and 2, further elucidate the above architecture.
\begin{equation}
z^{(i)} = F_{1}(x^{(i)};\theta1)    
\end{equation}
\begin{equation}
y^{(i)}_{pred} = F_{2}(z^{(i)};\theta2)    
\end{equation}
where $F_{1}$ and $F_{2}$ are the representations of the encoder and decoder with parameters $\theta_{1}$ and $\theta_{2}$, respectively. The weights and biases in the proposed architecture are optimized by minimizing the Smooth$L_{1}$ loss between  $y_{pred}^{(i)}$ and $y^{(i)}$. The Loss function L(X) is defined as: 
\begin{equation}
L(X) = \sum_{i=1}^{m} SmoothL_1(y_{diff})
\end{equation}
\begin{equation}
y_{diff} = y^{(i)} - y_{pred}^{(i)} 
\end{equation}
\begin{equation}
SmoothL_{1}(y_{diff}) =  \begin{cases} 0.5(y_{diff}^{2}), & \mbox{if } abs(y_{diff}) < 1 \mbox{} \\ abs(y_{diff}) - 0.5, &\mbox{otherwise,} \end{cases}
\end{equation}

\subsection{Model Architecture}
The architecture of the proposed model is adapted from the INcResU-Net network \cite{shankaranarayana2019fully} which is of immense importance in 2D medical image segmentation. The proposed architecture is shown in Fig. 1. The INcResU-Net, commonly referred to as the Encoder-Decoder, takes in an ECG record of length (1,5000) as input. The encoder downsamples the input through 8 layers of strided-convolutions. Downsampling by a factor of two is carried out by 1D convolution filters of stride 2 and kernel size 4. Convolution filters of stride 1 and kernel size 4 are used to adjust for odd feature sizes. The number of filters is increased by a factor of two until 1024, after which it is kept constant. In each layer, the 1D convolution is followed by a Batch Normalization layer, leaky ReLU activation(of slope 0.2) and Inception-Res block.    

\begin{figure}[t]
    \centering
    \includegraphics[width=0.8\linewidth]{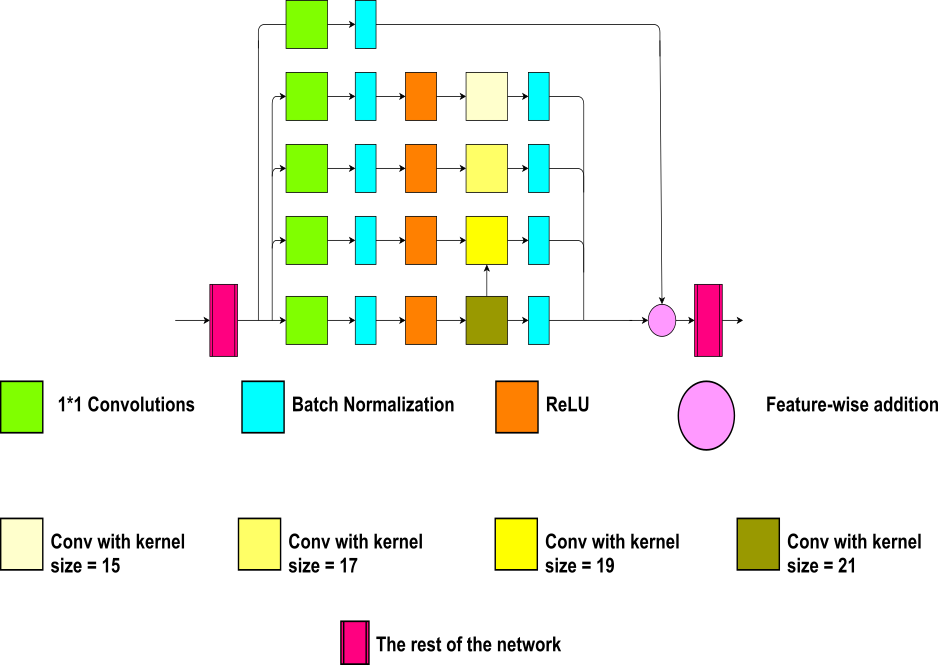}
    \caption{Inception-Res Block}
    \label{fig:Architecture}
\end{figure}

In the Inception-Resblock, a residual network which utilizes a skip connection mechanism is used. This helps to alleviate the vanishing gradient problem allowing the network to have depth. Furthermore, it also ensures quick convergence. Each residual network contains an Inception module (shown in Fig 2), that produces receptive fields of different sizes at the same layer. This is done at a reduced complexity by exploiting 1$\times$1 convolutions. The kernel sizes used in the Inception block, which were empirically arrived at, are 15,17,19 and 21. The output is upsampled by Transpose convolutions to extract a distance map of the underlying ECG waveform. Each feature map in the encoder is concatenated with the upsampled feature map of the corresponding size to retain structural coherence.  
\subsection{Dataset Description}
\subsubsection{CPSC}
The CPSC database \cite{gao2019open} contains 2000 ECG records, each of 10 seconds duration, sampled at 500 Hz. All the records are obtained from patients who had cardiovascular disease(CVD). Although the details of the acquisition and patient information are not available, this dataset is immensely important for beat detection, as it contains noisy ECG episodes. 
\subsubsection{MIT-BIH} The MIT-BIH dataset \cite{moody2001impact} consists of 48 half-hour excerpts of two-channel ambulatory ECG recordings, obtained from 47 subjects. This is a standard database that has high-quality ECG signals.
\subsubsection{MIT-BIH ST Change Database} The ST Change database \cite {albrecht1983st} includes 28 ECG recordings of varying lengths recorded during exercise stress tests. The presence of transient ST depression, ST elevation and motion artifacts, that arise during the exercise, make it an ideal candidate for evaluating the model.

\begin{table}[h]
\caption{Dataset split}
\label{Table}
\begin{center}
\begin{tabular}{||c c c c||} 
\hline
Dataset  & Total & Train & Test \\ [0.5ex] 
\hline\hline
CPSC & 2000 & 1936 & 64
\\
\hline
MIT-BIH & 	8640 & - & 8640
\\
\hline
MIT Exercise Stress Test & 4842 & - & 4842
\\
\hline
NSTDB & 1800 & - & 1800
\\
\hline
\end{tabular}
\end{center}
\end{table}

\subsubsection{NSTDB}The records in the NSTDB \cite{moody1984noise} were obtained by combining the intervals that contained baseline wander, muscle (EMG) artifact and electrode motion artifact with two clean records from the MIT-BIH database. SNR of range 24dB to -6 dB was thus obtained. This database allowed the study of model performance in the presence of noise.

\section {Experiments and Results}
\subsection {Implementation Details}
The parameters of the model were randomly initialized during training. Adam optimizer was used to optimize the SmoothL1 Loss between the obtained distance map and the original distance map. The network was trained for 500 epochs. An adaptive learning rate method was used, with the initial learning rate being set to 0.05. The learning rate was decreased by a factor of 10 for every 150 epochs. The model was implemented in Pytorch on a workstation housing an Nvidia GTX1080Ti 11GB GPU.\footnote{Code available at {https://github.com/acrarshin/RPNet}}
\subsection {Evaluation Method}

The model's performance is reported through three metrics. They include Precision, Recall, F1-score. Note that, the measurement of True Positives(TP), True Negatives(TN) and False Positives(FP) were made within a tolerance of 75 ms \cite{liu2018performance}.
\subsection{Results}
Three experiments are performed to highlight the robustness and versatility of the model. The experiments and their corresponding results are enumerated below. 

\subsubsection{\textbf{Comparison with three baselines on the CPSC dataset}} For this experiment, around 96 percent of the data from the CPSC dataset were randomly selected and was utilized for training. The rest was kept aside for evaluation of the model and for comparison with other beat detectors. A three fold cross validation was conducted to ascertain the best model. Comparisons are drawn against Hamilton \cite{hamilton2002open} and Christov \cite{christov2004real} detectors as they perform very well on the MIT-BIH database. Additionally, comparisons are carried out with Stationary Wavelet transform \cite{merah2015r} as it was shown to be highly robust to noise\footnote{Implemented using {https://github.com/berndporr/py-ecg-detectors}}. Table II highlights the results of these comparisons. All the traditional detectors were implemented as per the description given by the authors. 

\begin{table}[h]
\caption{Evaluation on CPSC}
\label{Table}
\begin{center}
\begin{tabular}{||c c c c||} 
\hline
Algorithm & Precision & Recall & F1-score \\ [0.5ex] 
\hline\hline
Hamilton & 0.7756 & 0.8621 & 0.8166
\\
\hline
Christov & 	0.7135 & 0.9085 & 0.7993
\\
\hline
SWT & 0.7791 & 0.8709 &	0.8224
\\
\hline
Ours & \textbf{0.9862} & \textbf{0.9812} & \textbf{0.98375}
\\
\hline
\end{tabular}
\end{center}
\end{table}

\subsubsection{\textbf{Establishing generalization capacity across three datasets}} The model was trained on all of the data in CPSC and was evaluated on the other databases. This is done to show that the proposed methodology is agnostic to changes in acquisition protocol acquisition device, sampling rate, and different kinds of artifacts.  All the records were split into ten-second windows. As these datasets were obtained at a sampling rate of 360 Hz, the windows are upsampled to 500 Hz and interpolated before the prediction is carried out. The obtained distance map is then, downsampled before post-processing. This ensures that the predictions of the beat locations align with the ground truth.
Table III summarizes the results obtained by the model on the three datasets.

\begin{table}[t]
\caption{Model's Performance on 3 datasets}
\label{Table}
\begin{center}
 \begin{tabular}{||c c c c||} 
 \hline
 Dataset & Precision & Recall & F1-score \\ [0.5ex] 
 \hline\hline
 MIT-BIH & 0.9944 & 0.9975 & 0.9965\\
 \hline
 MIT ST Change & 0.9972 & 0.9983 & 0.9978\\
 \hline
 NSTDB & 0.982 & 0.9451 & 0.9632\\
 \hline
\end{tabular}
\end{center}
\end{table}
\begin{table}[t]
\caption{Model's performance on NSTDB}
\label{Table}
\begin{center}
 \begin{tabular}{||c c c c||} 
 \hline
 SNR (db) & Precision & Recall & F1-Score \\ [0.5ex] 
 \hline\hline
 24 & 0.9986 & 0.9994 & 0.999
 \\
 \hline
 18 & 0.9979 & 0.9994 & 0.9986
 \\
 \hline
 12 & 0.9855 & 0.9986 & 0.992
 \\
 \hline
 6 & 0.9361 & 0.9859 & 0.9603
 \\
 \hline
 0 & 0.8228 & 0.9264 & 0.8715
 \\
 \hline
\end{tabular}
\end{center}
\end{table}
\subsubsection{\textbf{Analyzing the effect of noise on model performance}} The model was trained on the CPSC dataset and evaluated on the NSTDB dataset. This had involved evaluating records that have a particular SNR value. Six SNR values ranging from 24 dB to 0 dB were considered for this experiment. Records with an SNR of -6 dB were not considered for evaluation because of the absence of such noise levels during training. Note that, this particular result is an expansion of the evaluation on NSTDB as a whole. Table IV displays the model's predictive power at different levels of noise(different SNR's) on ECG records obtained from the NSTDB dataset.  

\begin{figure*}[t]
    \centering
    \includegraphics[width=13cm]{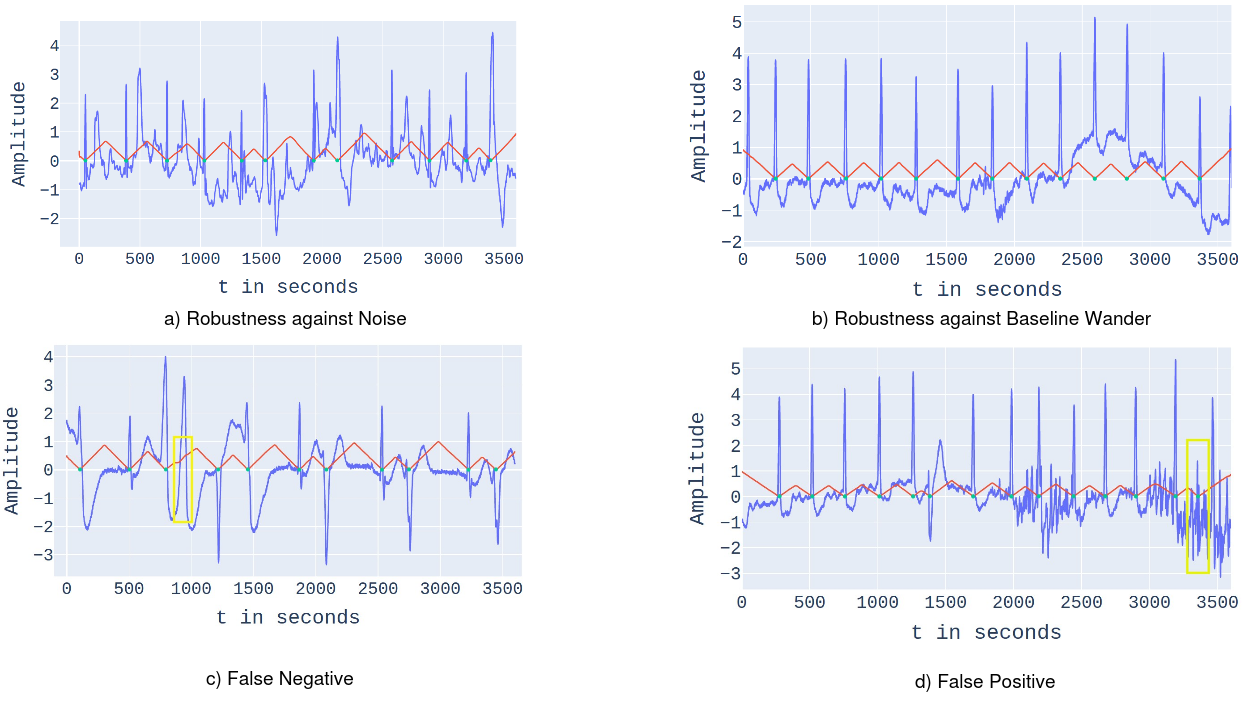}
    \caption{Qualitative analysis of ECG waveforms and predicted DT. Note that the ECG waveforms are shown in Blue and the predicted DT is shown in red. The ECG peak locations are highlighted in green. False positives and False negatives are highlighted in yellow.}
    \label{fig:Image}
\end{figure*}

\section{Discussion}
From Table II, it is evident that the model performs better on the CPSC dataset than all baselines, achieving an F1 score of 0.9837. The exceptional performance of the model is due to the choice of architecture, hyperparameters and a robust training methodology, which were arrived at empirically. The poor performance of the traditional baselines is due to the amount of artifacts present in CPSC.  

Table III shows the capacity of the model to generalize to other datasets. Note that there was no transfer learning before the evaluation of the proposed model on these datasets. This generalization is only possible because of the underlying morphological structure of the ECG. The performance on Excercise ST Database is surprisingly better when compared to that on the MIT-BIH dataset. This can be attributed to the relatively larger size of the MIT-BIH dataset, which is evident from Table I. As expected, the model performs the poorest on the NSTDB which is simulated with noise.   

A closer look at Table IV reveals a relatively poor predictive power at lower SNR levels, which is expected. This brings down the average F1 score, resulting in a lower score on the NSTDB dataset.  However, an F1 score of 0.8715 at an SNR of 0db is still acceptable, given that the signal can barely be differentiated from noise.

Some ECG records and their corresponding distance maps are shown in Figure 3. Fig.3a and Fig.3b show good approximations of the distance transform, where they highlight the robustness of the model against noise and baseline wandering respectively. The peak locations perfectly align with the ground truth annotation. Note that, it is hard for an untrained human eye to detect the R-peaks accurately in Fig.3a. Fig 3c and Fig 3d, show two failure cases from Exercise ST and MIT-BIH and are representative of a majority of the failure cases. In Fig 3c, it can be observed that the network does make an attempt to predict the peak only to be thrown off by the unusual width of the QRS complex. In Fig.3d, the ECG quality at the location of the False positive is visibly very poor.

\section{Conclusion}
The current work describes a novel application of the Unet architecture to obtain a distance map, from which the R-peaks are eventually extracted. Future work involves evaluation and comparison of computational complexity of the model in order to make it suitable for real time usage. Furthermore, it is imperative to explore robust quantization and model compression techniques to truncate the model and accelerate the inference process. 

\bibliographystyle{IEEEtran}
\bibliography{main}

\begin{thebibliography}{10}
\providecommand{\url}[1]{#1}
\csname url@samestyle\endcsname
\providecommand{\newblock}{\relax}
\providecommand{\bibinfo}[2]{#2}
\providecommand{\BIBentrySTDinterwordspacing}{\spaceskip=0pt\relax}
\providecommand{\BIBentryALTinterwordstretchfactor}{4}
\providecommand{\BIBentryALTinterwordspacing}{\spaceskip=\fontdimen2\font plus
\BIBentryALTinterwordstretchfactor\fontdimen3\font minus
  \fontdimen4\font\relax}
\providecommand{\BIBforeignlanguage}[2]{{%
\expandafter\ifx\csname l@#1\endcsname\relax
\typeout{** WARNING: IEEEtran.bst: No hyphenation pattern has been}%
\typeout{** loaded for the language `#1'. Using the pattern for}%
\typeout{** the default language instead.}%
\else
\language=\csname l@#1\endcsname
\fi
#2}}
\providecommand{\BIBdecl}{\relax}
\BIBdecl

\bibitem{manikandan2012novel}
M.~S. Manikandan and K.~Soman, ``A novel method for detecting r-peaks in
  electrocardiogram (ecg) signal,'' \emph{Biomedical Signal Processing and
  Control}, vol.~7, no.~2, pp. 118--128, 2012.

\bibitem{moody2001impact}
G.~B. Moody and R.~G. Mark, ``The impact of the mit-bih arrhythmia database,''
  \emph{IEEE Engineering in Medicine and Biology Magazine}, vol.~20, no.~3, pp.
  45--50, 2001.

\bibitem{liu2018performance}
F.~Liu, C.~Liu, X.~Jiang, Z.~Zhang, Y.~Zhang, J.~Li, and S.~Wei, ``Performance
  analysis of ten common qrs detectors on different ecg application cases,''
  \emph{Journal of healthcare engineering}, vol. 2018, 2018.

\bibitem{sannino2018deep}
G.~Sannino and G.~De~Pietro, ``A deep learning approach for ecg-based heartbeat
  classification for arrhythmia detection,'' \emph{Future Generation Computer
  Systems}, vol.~86, pp. 446--455, 2018.

\bibitem{andersen2019deep}
R.~S. Andersen, A.~Peimankar, and S.~Puthusserypady, ``A deep learning approach
  for real-time detection of atrial fibrillation,'' \emph{Expert Systems with
  Applications}, vol. 115, pp. 465--473, 2019.

\bibitem{murugesan2019psi}
B.~Murugesan, K.~Sarveswaran, S.~M. Shankaranarayana, K.~Ram, J.~Joseph, and
  M.~Sivaprakasam, ``Psi-net: Shape and boundary aware joint multi-task deep
  network for medical image segmentation,'' in \emph{2019 41st Annual
  International Conference of the IEEE Engineering in Medicine and Biology
  Society (EMBC)}.\hskip 1em plus 0.5em minus 0.4em\relax IEEE, 2019, pp.
  7223--7226.

\bibitem{shankaranarayana2019fully}
S.~M. Shankaranarayana, K.~Ram, K.~Mitra, and M.~Sivaprakasam, ``Fully
  convolutional networks for monocular retinal depth estimation and optic
  disc-cup segmentation,'' \emph{IEEE journal of biomedical and health
  informatics}, 2019.

\bibitem{gao2019open}
H.~Gao, C.~Liu, X.~Wang, L.~Zhao, Q.~Shen, E.~Ng, and J.~Li, ``An open-access
  ecg database for algorithm evaluation of qrs detection and heart rate
  estimation,'' \emph{Journal of Medical Imaging and Health Informatics},
  vol.~9, no.~9, pp. 1853--1858, 2019.

\bibitem{albrecht1983st}
P.~Albrecht, ``St segment characterization for long term automated ecg
  analysis,'' Ph.D. dissertation, Massachusetts Institute of Technology,
  Department of Electrical Engineering~…, 1983.

\bibitem{moody1984noise}
G.~B. Moody, W.~Muldrow, and R.~G. Mark, ``A noise stress test for arrhythmia
  detectors,'' \emph{Computers in cardiology}, vol.~11, no.~3, pp. 381--384,
  1984.

\bibitem{hamilton2002open}
P.~Hamilton, ``Open source ecg analysis,'' in \emph{Computers in
  cardiology}.\hskip 1em plus 0.5em minus 0.4em\relax IEEE, 2002, pp. 101--104.

\bibitem{christov2004real}
I.~I. Christov, ``Real time electrocardiogram qrs detection using combined
  adaptive threshold,'' \emph{Biomedical engineering online}, vol.~3, no.~1,
  p.~28, 2004.

\bibitem{merah2015r}
M.~Merah, T.~Abdelmalik, and B.~Larbi, ``R-peaks detection based on stationary
  wavelet transform,'' \emph{Computer methods and programs in biomedicine},
  vol. 121, no.~3, pp. 149--160, 2015.

\end{thebibliography}
\end{document}